\documentclass[twocolumn,floats,showpacs,amsmath,amssymb,prb]{revtex4}
\usepackage{graphicx}
\begin{document}

\title{Molecular hydrogen in graphite: A path-integral simulation}
\author{Carlos P. Herrero}
\author{Rafael Ram\'{\i}rez}
\affiliation{Instituto de Ciencia de Materiales,
         Consejo Superior de Investigaciones Cient\'{\i}ficas (CSIC),
         Campus de Cantoblanco, 28049 Madrid, Spain }
\date{\today}

\begin{abstract}
   Molecular hydrogen in the bulk of graphite has been studied 
by path-integral molecular dynamics simulations.
 Finite-temperature properties of H$_2$ molecules adsorbed between
graphite layers were analyzed in the temperature range from 300 to 900 K.
The interatomic interactions were modeled by a tight-binding potential
fitted to density-functional calculations. 
In the lowest-energy position, an H$_2$ molecule is found to be disposed
parallel to the sheets plane.
At finite temperatures, the molecule explores other orientations, but
its rotation is partially hindered by the adjacent graphite layers.
Vibrational frequencies were obtained from a linear-response approach,
based on correlations of atom displacements.
For the stretching vibration of the molecule, we find at 300 K a frequency 
$\omega_s$ = 3916 cm$^{-1}$, more than 100 cm$^{-1}$ lower than the 
frequency corresponding to an isolated H$_2$ molecule.
Isotope effects have been studied by considering also deuterium and
tritium molecules.
For D$_2$ in graphite we obtained $\omega_s$ = 2816 cm$^{-1}$, i.e.,
an isotopic ratio $\omega_s$(H)/$\omega_s$(D) = 1.39.
\end{abstract}



\maketitle

\section{Introduction}
In the last few years there has been a surge of interest
on carbon-based materials, specially on those composed of C
atoms displaying $sp^2$ hybridization. This group of materials 
includes carbon nanotubes and fullerenes, discovered in last decades, 
as well as graphene, found in recent years,\cite{ka07,ge07} and
the well-known graphite. 
These materials, apart from its interest in basic science, are
promising tools for diverse technological applications. Thus, 
carbon-based systems, in general, are considered as possible candidates 
for hydrogen storage.\cite{ko05,di01} 
Also, chemisorption on two-dimensional systems, such as graphene or 
graphite surfaces, is supposed to be important for catalytic 
processes.\cite{sl03}

The scientific and technological interest of hydrogen as an impurity 
in solids and on surfaces has existed for several decades. 
In principle, it seems to be one of the simplest impurities, but a 
clear understanding of its properties is not trivial because of its 
low mass, and needs the combination of advanced experimental and 
theoretical methods.\cite{es95,pe92}
In addition to its basic interest as an impurity, a remarkable
aspect of hydrogen in solids and surfaces is its capability of
passivating defects and forming complexes, facts that have been 
extensively studied for many years.\cite{es95,pe92,ze99}

 Experimental investigations of isolated hydrogen in graphite
turn out to be difficult because of the large sensitivity required
to detect this impurity, along with the presence of a large
amount of hydrogen trapped at the boundaries of graphite
crystallites.\cite{at02,wa07,wa04}
The stable configurations of hydrogen in the bulk of graphite have been
studied in several theoretical works,\cite{sh03,fe02,di03}
where special emphasis was laid upon both atomic and molecular forms of 
this impurity.
Moreover, theoretical techniques have been applied to investigate the 
diffusion, trapping, and recombination of hydrogen on a
graphite surface.\cite{fe03,fe04,sh05b,mo08}
In this respect, chemisorption of a single hydrogen atom on a graphene 
sheet has been studied by several authors using {\em ab-initio}
methods,\cite{sl03,du04,ya07,an08,bo08,ca09}
and their results show the appearance of a defect-induced magnetic moment,
along with a large structural distortion.\cite{ya07,bo08,ca09}

For the storage of hydrogen in graphite one should also consider
the presence of H$_2$ molecules in the graphite bulk, which are expected 
to be physisorbed in the interlayer space.\cite{di03,fe02,at02}
Here we will focus on isolated hydrogen molecules trapped between 
graphite sheets.
The importance of this problem is twofold: it is interesting
as a point defect in materials physics, for its relevance in 
the stability and diffusion of hydrogen in carbon-based solids,
and also H$_2$ in graphite is an example of a light molecule 
sitting and moving in a confined geometry, where quantum effects can
be nontrivial.

Earlier theoretical investigations of molecular hydrogen in solids
have focused on finding the lowest-energy position and stretching
frequency of the molecule, including sometimes  anharmonic effects
obtained from the calculated potential-energy 
surface,\cite{ok97,ok98,ho98,wa98,pr02}
and the quantum rotation of H$_2$ molecules.\cite{fo02,ho03}
Density-functional electronic-structure calculations in condensed matter 
are nowadays very reliable, but they usually deal with atomic nuclei as 
classical particles, so that typical quantum effects like zero-point 
vibrations are not directly included.
These effects can be taken into account by making use of harmonic or
quasiharmonic approximations, but are difficult to consider
when large anharmonicities are present, as may happen for
light impurities such as hydrogen.

The quantum character of the atomic nuclei can be taken into
account by using the path-integral molecular dynamics (or Monte Carlo) 
approach, which has been shown to be very useful in this respect.
A notable benefit of this procedure is that all nuclear degrees of
freedom can be quantized in an efficient and direct way, so that
both quantum and thermal fluctuations are directly included in the 
calculations.
Thus, molecular dynamics or Monte Carlo sampling applied to evaluate 
path integrals allows one to carry out quantitative and nonperturbative 
studies of anharmonic effects in many-body systems.\cite{gi88,ce95}

In the present paper, we use the path-integral molecular dynamics (PIMD) 
method to study hydrogen molecules adsorbed in the interlayer region of
graphite.
Particular emphasis was placed upon anharmonic effects in their quantum
dynamics at different temperatures.
We analyze the isotopic effect on structural and vibrational
properties of these molecules, by considering also molecular 
deuterium (D$_2$) and tritium (T$_2$). 
Path-integral methods similar to that employed in this work 
have been applied earlier to investigate hydrogen in metals\cite{gi88} 
and semiconductors,\cite{ra94,he95,mi98,he06,he07} as well as on 
surfaces.\cite{ma95b,he09a}
In relation to the behavior of molecular hydrogen in confined regions, 
H$_2$ has been studied inside carbon nanotubes by diffusion Monte 
Carlo.\cite{go01}
Also, path-integral simulation methods have been thoroughly applied 
to study condensed phases of hydrogen in molecular 
form.\cite{ch99,ka94,su97,ki00}

 The paper is organized as follows. In Sec.\,II, we describe the
computational method and the models employed in our calculations. 
Our results are presented in Sec.\,III, dealing with the 
spatial delocalization of H atoms, interatomic distance, 
vibrational frequencies, and kinetic energy. 
Sec.\,IV includes a summary of the main results.

\section{Computational Method}

\subsection{Path-integral molecular dynamics}

Our calculations are based on the path-integral formulation of 
statistical mechanics. In this formulation, the partition function is 
evaluated by a discretization of the density matrix along cyclic paths, 
made up of a finite number $L$ (Trotter number) of ``imaginary-time'' 
steps.\cite{fe72,kl90} 
In the implementation of this procedure to numerical simulations, 
such a discretization gives rise to the appearance of 
$L$ ``beads'' for each quantum particle. These beads can be treated in
the calculations as classical particles, since the partition function 
of the original quantum system is isomorph to that of a classical one. 
This isomorphism is obtained by replacing each quantum particle by a 
ring polymer consisting of $L$ classical particles, connected by harmonic 
springs.\cite{gi88,ce95}
In many-body problems, the configuration space can be adequately sampled by
molecular dynamics or Monte Carlo techniques. Here, we have used the
PIMD method, which was found to require less computer time for the
present problem.
We have employed effective algorithms for performing PIMD simulations 
in the canonical $NVT$ ensemble, as those described in the 
literature.\cite{ma96,tu02}

Calculations have been performed within the Born-Oppenheimer approximation, 
which allows us to define a $3N$-dimensional potential-energy surface 
for the motion of the atomic nuclei.
An important issue in this type of calculations is the proper description 
of interatomic interactions, which should be as realistic as possible.
Since effective classical potentials present many limitations
to reproduce the many-body energy surface, one should
resort to self-consistent quantum-mechanical methods.
However, density functional (DF) or Hartree-Fock-based self-consistent 
potentials
require computing resources that would appreciably restrict the size of
our simulation cell and/or the number of simulation steps.
We found a reasonable compromise by obtaining the Born-Oppenheimer 
surface from a tight-binding (TB) effective Hamiltonian, derived from 
DF calculations.\cite{po95}
The capability of TB methods to simulate different properties of
solids and molecules has been reviewed by Goringe {\em et
al.}\cite{go97}
In particular, the ability of our DF-TB potential to predict 
frequencies of C--H vibrations in molecular systems was shown
in Refs.~\onlinecite{lo03,bo01}.
We have employed earlier this TB Hamiltonian to describe
hydrogen-carbon interactions in diamond\cite{he06,he07} and
graphene.\cite{he09a}
The TB energy consists of two parts; one of them is the sum of
energies of occupied one-electron states, and the other
corresponds to a pairwise interatomic potential.\cite{po95}
Since a reliable description of the hydrogen molecule is essential
for our purposes, particular attention was put on the H-H pair potential,
which has been taken as in our earlier study of molecular hydrogen in
silicon.\cite{he09b}  This pair potential reproduces
the main features of known effective interatomic potentials for H$_2$,
such as the Morse potential.\cite{ra01}

  Simulations were carried out on a graphite supercell containing 
64 C atoms and one hydrogen molecule (H$_2$, D$_2$, or T$_2$), and 
periodic boundary conditions were assumed.
The simulation cell includes two graphite sheets, each one being a
$4\times4$ graphene supercell of size $4 a$ = 9.84 \AA.
An $AB$ layer stacking was considered, so that both sheets are disposed
in such a way that the center of each hexagonal ring of one of them
lies over a C atom of the adjacent sheet.
To keep this type of stacking along a simulation run, 
avoiding diffusion of the graphite layers, the
center-of-gravity of each layer was not allowed to move on the
layer plane, which will be referred in the sequel as the
$(x, y)$ plane. 
The average distance between sheets is a half of the supercell 
parameter along the $z$ axis (perpendicular to the graphite layers), 
and was taken to be 3.35 \AA.
For the reciprocal-space sampling we have used only the $\Gamma$
point (${\bf k} = 0$), since the effect of employing a larger ${\bf k}$
set is a nearly constant shift in the total energy, with little
influence on the energy differences between different atomic
configurations.
The influence of the cell size on the results of the simulations has
been checked by considering graphite supercells including up to 144 C
atoms (a $6 \times 6$ supercell). The results found for $4 \times 4$,
$5 \times 5$, and $6 \times 6$ supercells coincided within
statistical error bars. In particular, we checked the kinetic 
energy of the H$_2$ molecule (error bar of $\pm 3$ meV) and the 
mean H--H distance (error bar of $\pm 2 \times 10^{-4}$ \AA).
Also, including more graphite layers in the simulation cell does not
affect the results of the PIMD simulations.

Sampling of the configuration space has been carried out at temperatures 
between 300 and 900 K.  For comparison, we also 
carried out PIMD simulations of pure graphite, as well as simulations of
hydrogen  molecules between rigid graphite sheets (in which the 
C atoms are kept fixed on their unrelaxed positions; see Sect.~II.B for
a precise definition of this approach). Moreover, some simulations were 
performed in the classical limit, which is obtained in our context by
setting the Trotter number $L = 1$.
The electronic-structure calculations were performed without considering a
temperature-dependent Fermi filling of the electronic states, which is
reasonable for the temperature range under consideration.
 For a given temperature, a typical simulation run consisted of 
$2 \times 10^4$ PIMD steps for system equilibration, followed by 
$10^6$ steps for the calculation of ensemble average properties.
To keep roughly a constant precision in the PIMD results
at different temperatures, the Trotter number was scaled with the 
inverse temperature ($L \propto 1/T$), so that $L T$ = 18000 K, 
which translates into $L$ = 60 for  $T$ = 300 K.
Quantum exchange effects between hydrogen nuclei were not taken
into account, as they are negligible at the temperatures considered here, 
and both atomic nuclei in a molecule were treated as distinguishable
particles.

The simulations were performed by using a staging transformation
for the bead coordinates.
The canonical ensemble was generated by coupling chains of four 
Nos\'e-Hoover thermostats to each degree of freedom.\cite{tu98}
To integrate the equations of motion, we employed
a reversible reference-system propagator algorithm (RESPA), which allows
one to define different time steps for the integration of fast and slow
degrees of freedom.\cite{ma96} 
The time step $\Delta t$ associated to the calculation of DF-TB forces
was taken in the range
between 0.1 and 0.3 fs, which was found to be appropriate for the
interactions, atomic masses, and temperatures considered here.
For the evolution of the fast dynamical variables, associated to the
thermostats and harmonic bead interactions, we used a smaller
time step $\delta t = \Delta t/4$.
Note that for H$_2$ in graphite at 300 K, a simulation run consisting of 
$10^6$ PIMD steps needs the calculation of energy and forces with the TB 
code for $6 \times 10^7$ configurations, which required the use of 
parallel computing.

\subsection{Path centroid delocalization}

We now define some spatial properties of the particle paths that will
be used in the analysis of the simulation results.
The center-of-gravity (centroid) 
of the quantum paths of a given particle is calculated as
\begin{equation}
   \overline{\bf r} = \frac{1}{L} \sum_{i=1}^L {\bf r}_i  \, ,
\label{centr}
\end{equation}
${\bf r}_i$ being the position of bead $i$ in the associated ring
polymer.

The mean-square displacement of a quantum particle along a PIMD
simulation run is then given by:
\begin{equation}
\Delta^2_r =  \frac{1}{L} \left< \sum_{i=1}^L 
           ({\bf r}_i - \left< \overline{\bf r} \right>)^2
           \right>    \, ,
\label{delta2}
\end{equation}
where $\langle ... \rangle$ indicates a thermal average at temperature $T$.
After a direct transformation, one can write $\Delta^2_r$ as
\begin{equation}
 \Delta^2_r = Q^2_r + C^2_r  \, ,
\label{delta2b}
\end{equation}
with
\begin{equation}
 Q^2_r = \frac{1}{L} \left< \sum_{i=1}^L 
             ({\bf r}_i - \overline{\bf r})^2 \right>    \, ,
\end{equation}
and
\begin{equation}
  C^2_r = \left< (\overline{\bf r} - 
                \left< \overline{\bf r} \right>)^2 \right> 
      = \left< \overline{\bf r}^2 \right> -
             \left< \overline{\bf r} \right>^2     \, .
\label{cr2}
\end{equation}
The first term, $Q^2_r$, is the mean-square ``radius-of-gyration''
of the ring polymers associated to the quantum particle 
under consideration, and gives the average spatial
extension of the paths.\cite{gi88}
The second term on the r.h.s. of Eq.~(\ref{delta2b}), $C^2_r$, 
is the mean-square displacement of the path centroid.
This term is the only one remaining in the high-temperature (classical) 
limit, since then each path collapses onto a single point and the 
radius-of-gyration vanishes.
In cases where the anharmonicity is not very large, the spatial
distribution of the centroid $\overline{\bf r}$ is similar to that 
of a classical 
particle moving in the same potential, and $C^2_r$ can be considered
as a kind of semiclassical delocalization. 

As indicated above, results of our PIMD simulations for H$_2$ in 
graphite will be compared with those obtained for rigid graphite layers. 
This means that in the latter case the centroids of the quantum paths 
corresponding to carbon atoms are kept fixed on their ideal atomic
positions, so that no relaxation of the host atoms is allowed in the
presence of the hydrogen molecule. This restriction allows however 
paths of the C atoms to be delocalized around their ideal sites,
i.e. in this approach one has $C^2_r = 0$ and $Q^2_r > 0$ for the
carbon atoms.

\subsection{Anharmonic vibrational frequencies}

  Vibrational frequencies are often employed as fingerprints of 
impurities in solids, revealing information on the position that they 
occupy and on their interactions with the nearby hosts atoms.
A traditional approach for calculating vibrational frequencies of 
impurities consists in obtaining the eigenvalues of the dynamical matrix
associated to the atoms in the simulation cell, which yields the 
frequencies in a harmonic approximation. 
However, for light impurities the anharmonicity can be large, and 
the harmonic frequencies are only a first (maybe poor) approximation.

Anharmonic frequencies of vibrational modes will be calculated here 
by using a method based on the linear response (LR) of the system to 
vanishingly small forces applied on the atomic nuclei.
In the context of path-integral simulations,
this approach has been shown to represent a significant 
improvement with respect to a standard harmonic approximation.\cite{ra01}
In particular, the vibrational frequency of the H$_2$ stretching mode 
is derived by the LR method as    
\begin{equation}
\omega_s = \left( \frac{k_B T}{\mu C^2_d} \right)^{\frac12}  \;,
\end{equation}
where $k_B$ is Boltzmann's constant, $\mu$ is the reduced mass of the 
H$_2$ molecule, and 
$C^2_d$ is the mean-square displacement of the H-H distance, $d$, 
that is obtained by a relation analogous to Eq. (\ref{cr2}), after
substitution of the particle coordinate $\bf{r}$ 
by the interatomic distance $d$.
Details on this method and discussions of its capability for predicting
vibrational frequencies of molecules and solids are given
elsewhere.\cite{ra01,ra02,lo03,ra05}

\section{Results}

\subsection{Atomic delocalization}

For an H$_2$ molecule we find a minimum-energy position at an
interstitial site between a carbon atom in a graphite sheet and an
hexagonal ring in an adjacent sheet. At this position, the preferred
orientation of the molecule is parallel to the graphite planes,
in agreement with earlier calculations based on density-functional
theory.\cite{di03}
At finite temperatures the molecule will explore other
positions and orientations with respect to the graphite layers.
In particular, the molecule can be tilted, forming an angle $\varphi$ 
with the $(x,y)$ plane.

\begin{figure}
\vspace{-2.0cm}
\hspace{-0.5cm}
\includegraphics[width= 9cm]{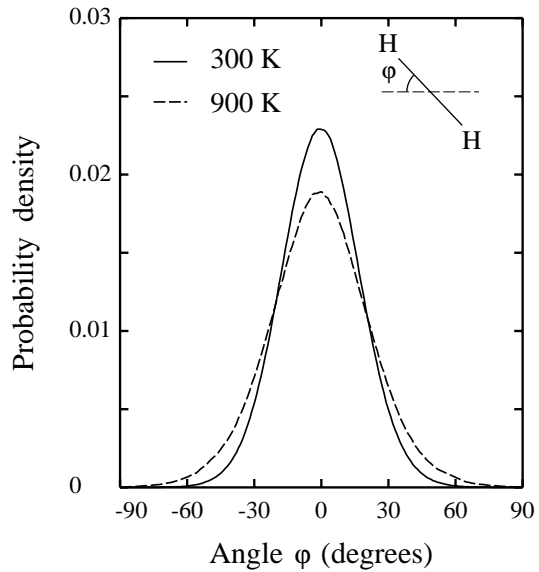}
\vspace{-2.5cm}
\caption{
Probability distribution of the angle $\varphi$ between the
H--H direction and the graphite sheets, as derived from PIMD
simulations at two temperatures: 300 K (solid
line) and 900 K (dashed line).
}
\label{f1}
\end{figure}

In Fig.~1 we present the probability distribution of the angle
$\varphi$,  as derived from our PIMD simulations at two temperatures:
300 K (solid line) and 900 K (dashed line).
This distribution has a maximum at $\varphi = 0$ (H--H parallel to the
layers), and vanishes
for H--H perpendicular to the sheet plane ($\varphi = \pm 90^o$).
As temperature increases, the probability distribution broadens
slightly, but it remains as a peak centered at $\varphi = 0$.
However, we find that the molecule is free to rotate in the
$(x, y)$ plane.

\begin{figure}
\vspace{-2.0cm}
\hspace{-0.5cm}
\includegraphics[width= 9cm]{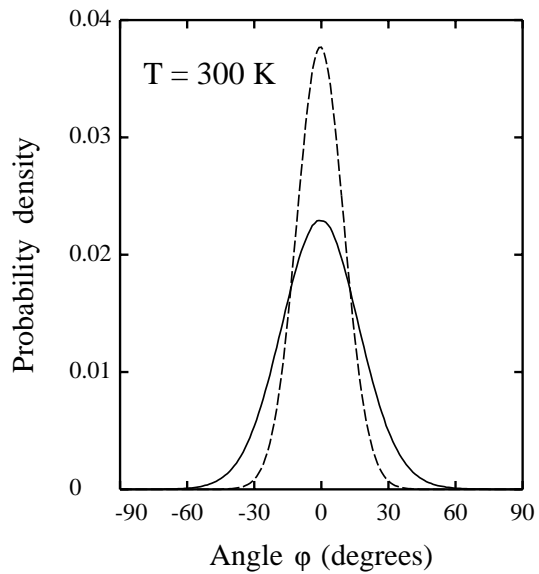}
\vspace{-2.5cm}
\caption{
Probability distribution of the angle $\varphi$ between the
H--H direction and the graphite sheets, as derived from PIMD
simulations at 300 K. Shown are results obtained for mobile
C atoms (no restrictions, solid line), as well as from a
simulation in which the C atoms where kept fixed on their unrelaxed
positions (fixed centroids, dashed line).
}
\label{f2}
\end{figure}

Even though there is no direct bond between molecular hydrogen and the
graphite layers, the later relax slightly in the presence of the
H$_2$ molecules, so that they follow the H$_2$ motion in the interlayer
space. This means that the molecules are more mobile in the presence of
flexible layers than in the case of stiff graphite sheets, in which the
C atoms are fixed on their unrelaxed (ideal) positions.
This can be visualized by looking at the distribution of the angle
$\varphi$ in both cases at a given temperature. 
Thus, in Fig.~2 we display this probability distribution at $T$ = 300 K.
The dashed line corresponds to rigid graphite sheets, whereas the solid
one was obtained for flexible sheets (mobile C atoms).
As expected, the distribution of the angle $\varphi$ is broader for
flexible sheets, since in this case the H$_2$ molecule can adopt
configurations that are inaccessible in the presence of rigid graphite 
layers.
We have also calculated the same probability distribution for D$_2$
and T$_2$ for flexible sheets, and at 300 K it turns out to be similar 
to that shown in Fig.~2 for H$_2$ (solid line), but slightly narrower.

\begin{figure}
\vspace{-2.0cm}
\hspace{-0.5cm}
\includegraphics[width= 9cm]{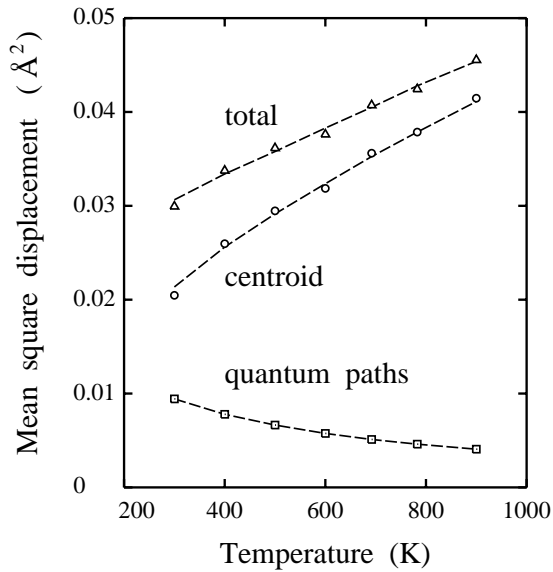}
\vspace{-2.5cm}
\caption{
Spatial delocalization of atomic nuclei (protons) in H$_2$ along the
$z$ coordinate (perpendicular to the sheet plane), as a function of
temperature.
Circles represent the mean-square displacement of the centroid of the
quantum paths, $C_z^2$, and squares correspond to the mean-square
radius-of-gyration of the paths, $Q_z^2$. The total delocalization,
$\Delta_z^2$, is shown by triangles.
Error bars of the total and centroid delocalization are in the order
of the symbol size, whereas those of the radius-of-gyration are less
than the symbol size.
Dashed lines are guides to the eye.
}
\label{f3}
\end{figure}

We now turn to study the spatial delocalization of each atomic
nucleus in hydrogen  molecules, that is expected 
to include a nonnegligible quantum contribution.
For our problem of H$_2$ in graphite, we have calculated separately
both terms giving the atomic delocalization in Eq.~(\ref{delta2b}),
for each atom in the molecule.
For a given temperature, the term $C^2_r$ does not converge to 
a well-defined value along a PIMD simulation, due to the onset of
molecular diffusion in the interlayer space.
However, its component $C_z^2$ along the $z$ axis is
an equilibrium property of the molecule, as in fact it cannot diffuse 
across graphite layers.
In Fig.~3 we display the values of $Q_z^2$ (spreading of the quantum
paths, squares) and $C_z^2$ (centroid delocalization, circles),
as derived from our PIMD simulations for the H$_2$ molecule at several 
temperatures. 
The total spatial delocalization along the $z$ coordinate,
$\Delta_z^2$, is shown as triangles.
In this plot, one observes that $C_z^2$ is larger than 
$Q_z^2$ in the whole temperature range considered.
From the spatial delocalization $Q_z^2$ shown in Fig.~3, one can
estimate an effective frequency for hydrogen motion in the $z$
direction. In fact, in a harmonic approximation $Q_z^2$ can be
expressed analytically as a function of frequency and
temperature,\cite{gi88,ra93} and comparing the delocalization
expected for different frequencies with that given by our PIMD
simulations, we found an effective frequency $\omega_z$ of about
1200 cm$^{-1}$. 

For the spreading of the quantum paths of each H atom we obtain at
room temperature $Q_z^2 = 9.4 \times 10^{-3}$ \AA$^2$, and it decreases 
as temperature is raised.
It is interesting to compare this value with that found for $Q_z^2$
in the case of unrelaxed graphite layers. When the C atoms are fixed
on their ideal positions, we find at 300 K, 
$Q_z^2 = 5.9  \times 10^{-3}$ \AA$^2$, much lower than that found when
the C atoms are allowed to relax in the presence of the hydrogen molecule.
It is also interesting to compare these values of the quantum
delocalization in the direction perpendicular to the graphite sheets,
with that on the $(x, y)$ plane.
Our simulations yield $Q_x^2 = Q_y^2 = 1.07 \times 10^{-2}$ \AA$^2$
and $9.9 \times 10^{-3}$ \AA$^2$, for free and fixed C atoms,
respectively, indicating that the quantum motion of hydrogen on the 
$(x, y)$ plane is only slightly affected by motion of the carbon atoms.
In fact, diffusion of the H$_2$ molecules in the interlayer space 
occurs basically by classical jumps, as described elsewhere.\cite{he10} 
However, the quantum motion in the $z$ direction is markedly affected
by the relaxation of the C atoms, that contributes to enhance $Q_z^2$ 
by a factor of 1.6.
In other words, this increase in the quantum delocalization is 
associated to a decrease in frequency (softening) of the vibrational 
modes of the H$_2$ molecule when full motion of the C atoms is taken 
into account.
These modes correspond to a displacement of the whole molecule along the
$z$ axis, and to the frustrated molecular rotation, with changes in the 
angle $\varphi$ shown in Fig.~1. 
 We note that the quantum paths have an average extension of 
$\sim$ 0.1 \AA\ at 300 K, 
much smaller than the H-H distance, thus justifying the 
neglect of quantum exchange between protons. 
 
For the D$_2$ molecule we obtain at 300 K, 
$Q_z^2 = 5.4 \times 10^{-3}$ \AA$^2$
in the case of free motion of all atoms in the simulation cell. 
Comparing with the H$_2$ molecule, we have 
$Q_z^2$(H)/$Q_z^2$(D) = 1.7, clearly higher than the low-temperature
limit in a harmonic approximation, given by a ratio of $\sqrt{2}$.
Note that in the high-temperature limit $Q_z^2$ goes to zero,
but the ratio $Q_z^2$(H)/$Q_z^2$(D) converges to the inverse mass 
ratio,\cite{gi88,ra93} in this case $m_{\rm D}/m_{\rm H}$ = 2.
For T$_2$ we found at 300 K, $Q_z^2 = 3.8 \times 10^{-3}$ \AA$^2$,
so that $Q_z^2$(H)/$Q_z^2$(T) = 2.5, also between a ratio of 
$\sqrt{3}$ expected at low temperature in a harmonic approach,
and the high-temperature limit given by $m_{\rm T}/m_{\rm H}$ = 3.

\subsection{Interatomic distance}

We first present results for classical calculations at
zero temperature, where the atoms are treated as point-like
particles without spatial delocalization.
The interatomic potential employed here gives reliable results 
for molecular hydrogen in vacuo (an isolated H$_2$ molecule).
In particular, the lowest-energy molecular configuration corresponds 
to a distance $R_0$ between hydrogen atoms of 0.741 \AA.
At this distance we obtain for H$_2$ in a harmonic approximation
a stretching frequency of 4397 cm$^{-1}$.

\begin{figure}
\vspace{-2.0cm}
\hspace{-0.5cm}
\includegraphics[width= 9cm]{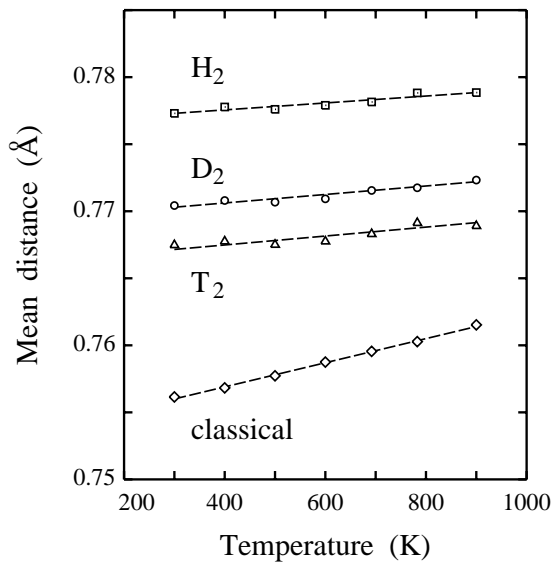}
\vspace{-2.5cm}
\caption{
Mean distance between atoms in a hydrogen  molecule in graphite,
as derived from PIMD simulations for H$_2$ (squares), D$_2$
(circles), and T$_2$ (triangles).
 Results obtained from classical molecular dynamics simulations
are also shown for comparison (diamonds).
Error bars are in the order of the symbol size.
Dashed lines are linear fits to the data points.
}
\label{f4}
\end{figure}

The interatomic distance between hydrogen
atoms increases when the molecule is introduced from the gas phase into 
the graphite bulk, due to an attractive interaction between H and the
nearby C atoms.
For the minimum-energy distance we found in this case $R_0$ = 0.753~\AA,
which is similar to that found for the H$_2$ molecule in the bulk of
semiconductor materials with the same interatomic potential.\cite{he09b}
This interatomic distance is expected to rise for increasing
temperature. In fact, in a classical approximation we obtained
$R$ = 0.756 \AA\ and 0.761 \AA\ at 300 K and 900 K, respectively.
These classical results are presented in Fig.~4 as diamonds, and
display a linear temperature dependence with slope
$d R/d T = 8.8 \times 10^{-6}$~\AA/K.

PIMD simulations can be also employed to study the temperature dependence
of the mean interatomic distance $R$ in a quantum model. The molecular 
expansion with respect to the lowest-energy classical geometry is due 
to a combination of anharmonicity in the stretching vibration of the
H$_2$ molecule and a centrifugal contribution caused by molecular
rotation. 
At 300 K we find for H$_2$ a mean interatomic distance $R$ = 0.777 \AA,
to be compared with $R$ = 0.756 \AA\ in the classical limit at the same
temperature. This means that the interatomic distance of H$_2$ in
graphite increases by 0.02 \AA\ (about a 3\%) when quantum effects 
are considered.
For the molecules D$_2$ and T$_2$, one expects smaller interatomic 
distances due to their larger mass and smaller vibrational amplitudes. 
In fact, at 300 K we found for D$_2$, $R$ = 0.770 \AA\ 
(i.e., $7 \times 10^{-3}$ \AA\ less than for H$_2$), and for 
T$_2$, $R$ = 0.767 \AA.   
In Fig.~4 we present the temperature dependence of the mean distance for
H$_2$ (squares), D$_2$ (circles), and T$_2$ (triangles), as derived 
from our PIMD simulations for full quantum motion of molecular hydrogen 
and host atoms. 
For D$_2$ and T$_2$ we find a slope $d R/d T = 3.1 \times 10^{-6}$ \AA/K
and $2.8 \times 10^{-6}$ \AA/K, respectively, close to
the value obtained for H$_2$: $d R/d T = 2.6 \times 10^{-6}$ \AA/K.
Note that these changes of the interatomic distance derived from the
PIMD simulations are much smaller than that found in the classical
limit: $d R/d T = 8.8 \times 10^{-6}$ \AA/K.

It is interesting to compare these changes in the mean distance $R$
with those corresponding to molecular hydrogen in the gas phase. 
With this purpose we carried out some PIMD simulations of an 
isolated hydrogen molecule with the same interatomic potential at
several temperatures. 
These simulations yielded an increase in $R$ as temperature is raised,
given by $d R/d T = 7.5 \times 10^{-6}$ \AA/K, a value three times 
larger than that found for H$_2$ in graphite.

\begin{figure}
\vspace{-2.0cm}
\hspace{-0.5cm}
\includegraphics[width= 9cm]{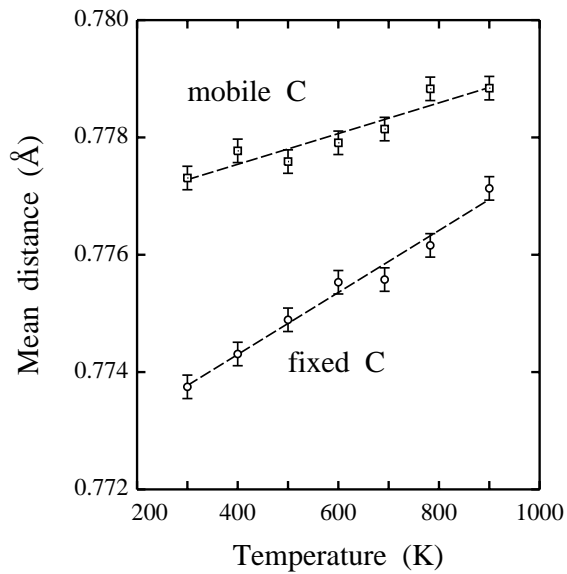}
\vspace{-2.5cm}
\caption{
Mean interatomic distance for H$_2$ molecules in graphite,
as a function of temperature.
Squares indicate results derived from PIMD simulations
in which all H and C atoms are mobile, whereas circles correspond
to simulations in which all C atoms where kept fixed on their unrelaxed
positions (fixed centroids).
Dashed lines are linear fits to the data points.
}
\label{f5}
\end{figure}

It is also interesting to analyze the effect of the motion of the
graphite layers on the H--H distance. We have calculated this distance
in PIMD simulations in which the C atoms are kept fixed on their unrelaxed
sites. The results of the interatomic distance are shown in Fig.~5 as
circles. We find that the H--H distance is in this case smaller than
that obtained for full quantum motion of all the atoms in the simulation
cell. In fact, at 300 K the distance $R$ decreases by about 
$4 \times 10^{-3}$ \AA. However, the slope $d R/ dT$ for the 
fixed-lattice model is larger than in the case of mobile C atoms.
In fact, for fixed C atoms we find $d R/d T = 5.3 \times 10^{-6}$ \AA/K,
to be compared with a slope of $2.6 \times 10^{-6}$ \AA/K obtained for
flexible graphite sheets.

From the zero-temperature classical calculations, we found that the
H$_2$ molecule is expanded when it is introduced into the graphite bulk,
as a consequence of the attractive C--H interaction. At finite
temperatures, the H--H distance is also controlled by the centrifugal
effect caused by molecule rotation. In fact, a 3D rotation is partially 
frustrated in the interlayer region, as shown above.
In this respect, molecular rotation is favored by the relaxation of
graphite sheets, thus yielding a larger centrifugal contribution to the
molecule expansion than for rigid sheets.

The average interatomic distance allows us to estimate a moment of
inertia for the molecule, and then the wave-number difference 
$\omega_{01}$ between $J = 0$ and $J = 1$ rotational levels. 
This gives for H$_2$,  $\omega_{01} \approx 110$ cm$^{-1}$, somewhat 
lower than that known for the free molecule in vacuo\cite{st57}
($\omega_{01} = 118.6$ cm$^{-1}$). For H$_2$ in graphite, however, 
the $J = 1$ level will be split due to the hindered motion of the 
molecule for H--H perpendicular to the graphite sheets. Unfortunately, 
the magnitude of this splitting is not accessible from our PIMD 
simulations.

\subsection{Stretching frequency} 

The stretching frequency of H$_2$ is an important fingerprint of 
the molecule, that can be used to detect and characterize
this impurity in solids.
We first note that the lowest-energy configuration of an isolated
H$_2$ molecule allows us to predict a stretching frequency of 
4397 cm$^{-1}$ in a harmonic approximation.
From path-integral simulations combined with the LR approach presented
above in Sect.~II.C, we obtain for a single H$_2$ molecule at 300 K
a frequency
$\omega_s = 4055 \pm 5$ cm$^{-1}$, i.e., the anharmonic shift amounts 
to more than 300 cm$^{-1}$. Note that in the vibrational frequencies
derived from the LR procedure, the error bars are due to the 
statistical uncertainty associated to the PIMD simulations. 

\begin{figure}
\vspace{-2.0cm}
\hspace{-0.5cm}
\includegraphics[width= 9cm]{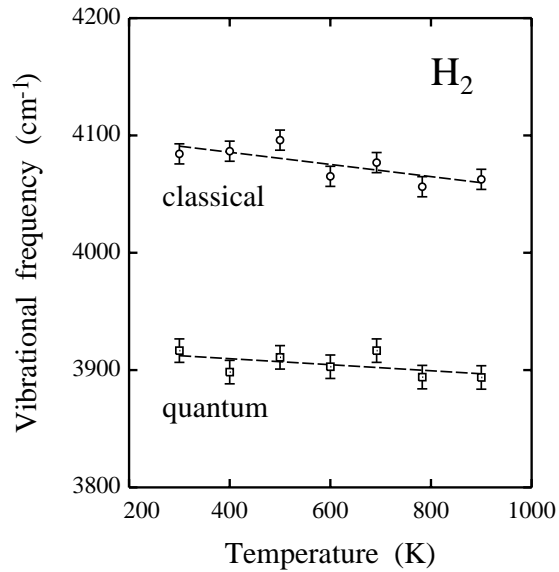}
\vspace{-2.5cm}
\caption{
Frequency of the stretching vibration of the H$_2$ molecule in graphite
as a function of temperature.
Symbols represent results derived from PIMD simulations in two
different approaches:
squares, full quantum motion of H and C atoms;
circles, classical motion of H and C atoms.
Dashed lines are linear fits to the data points.
Error bars correspond to the statistical uncertainty in the molecular
dynamics simulations.
}
\label{f6}
\end{figure}

The stretching frequency $\omega_s$ is further reduced when the H$_2$ 
molecule is inserted between the graphite sheets. In fact, from our PIMD
simulations at 300 K we found for H$_2$ a stretching frequency 
$\omega_s$ = $3916 \pm 8$ cm$^{-1}$.
This frequency is found to decrease slightly as the temperature is 
raised, as shown in Fig.~6 (squares). 
A linear fit to the data points gives a slope
$d \omega_s / d T = - 0.03$  cm$^{-1}$/K.

The quantum treatment of atomic nuclei in molecular
dynamics simulations is decisive to give a reliable description of
the vibrational frequencies of light atoms like hydrogen.
In fact, we have applied the LR method to calculate the stretching
frequency $\omega_s$ from classical simulations. At 300 K we
found for H$_2$ in graphite a frequency 
$\omega_s = 4082 \pm 7$ cm$^{-1}$ 
(for full motion of interstitial hydrogen and host atoms), 
about 160 cm$^{-1}$ higher than that found in the full quantum 
simulations ($\omega_s$ = 3916 cm$^{-1}$).
In Fig.~6 we have plotted the frequency $\omega_s$ derived from
the classical molecular dynamics simulations in the temperature
range from 300 to 900 K.
The overestimation of vibrational frequencies in a classical approach,
in comparison with the quantum results is usual in this kind of
simulations,\cite{he09b} since the classical calculations tend to 
yield results
much closer to the harmonic approximation, which gives in general
frequencies higher than the anharmonic ones (as is the case here).

As presented above when discussing the interatomic distance 
in the hydrogen molecule, there are two main factors controlling the
stretching frequency of the molecule in the graphite bulk. The first 
one is the interaction with the graphite sheets, which tends to
enlarge the H--H distance, with a concomitant decrease in the 
frequency $\omega_s$. This is the main factor contributing to the
reduction observed when comparing results of an isolated molecule and
a molecule in the interlayer region.
The second factor is the molecular rotation, which is partially
hindered between the graphite sheets, but in general causes a 
decrease in the vibrational frequency due to ro-vibrational
coupling. All together, we find that $\omega_s$ decreases only slightly
as $T$ is raised. This contrasts with the results obtained for 
the stretching frequency of H$_2$ in the interstitial space of
silicon from PIMD simulations similar to those presented
here.\cite{he09b}
In that case, the molecule is free to rotate in the silicon bulk,
and $\omega_s$ is found to decrease about eight times faster than
for H$_2$ in graphite. 

For the D$_2$ molecule in graphite we find at 300 K a stretching 
frequency $\omega_s = 2816 \pm 5$ cm$^{-1}$. 
For increasing temperature, we obtained a trend similar to that found
for H$_2$, with a linear decrease in $\omega_s$. 
For the isotopic shift we found a rather constant ratio
between the stretching frequencies of H$_2$ and D$_2$, that amounts to
1.39, slightly smaller than the ratio expected in a harmonic
approximation: $\omega_s$(H)/$\omega_s$(D) = 1.41.
Experimentally, a ratio of 1.39 has been observed for the frequencies 
of these molecules in the gas phase.\cite{st57}
For comparison, we mention that in a classical simulation of D$_2$
at 300 K we found a stretching frequency $\omega_s = 2896 \pm 5$ cm$^{-1}$,
which yields an isotopic ratio of 1.41, as in a harmonic approach.
From PIMD simulations of T$_2$ in graphite we obtained at 300 K, 
$\omega_s = 2324 \pm 5$ cm$^{-1}$, so that
$\omega_s$(H)/$\omega_s$(T) = 1.69, slightly lower than the harmonic
expectancy of 1.73.

\subsection{Kinetic energy}

Path integral simulations allow one to obtain the kinetic energy $E_k$ 
of the quantum particles under consideration, which is basically related 
to the spread of the quantum paths. In fact, for a particle at a
given temperature, the larger the mean-square radius-of-gyration of 
the paths, $Q_r^2$, the smaller the kinetic energy.
Here we have calculated $E_k$ by using the so-called virial
estimator, which has an associated statistical uncertainty lower
than the potential energy of the system.\cite{he82,tu98}

\begin{figure}
\vspace{-2.0cm}
\hspace{-0.5cm}
\includegraphics[width= 9cm]{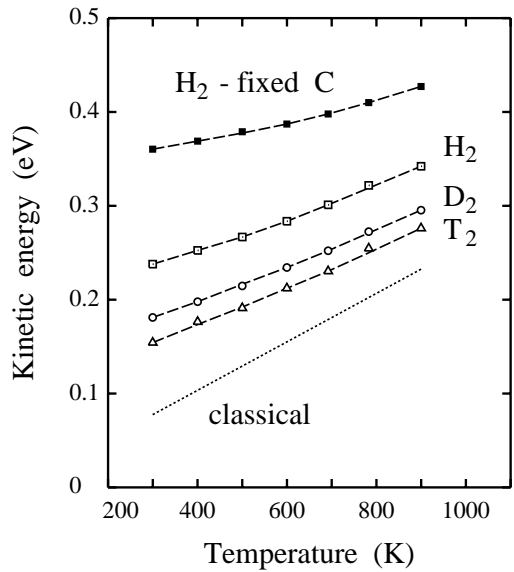}
\vspace{-2.5cm}
\caption{
Temperature dependence of the kinetic energy of molecular hydrogen in
graphite, as derived from PIMD simulations.
Open symbols indicate results derived from simulations with full
quantum motion of all atoms in the cell: squares for H$_2$,  circles for
D$_2$, and triangles for T$_2$. For comparison we also present results
for H$_2$ with fixed C atoms (filled squares). 
Error bars are on the order of the symbol size.
Dashed lines are guides to the eye.
The dotted line corresponds to the classical limit with six
degrees of freedom ($E_k^{\text cl} = 3 k_B T$).
}
\label{f7}
\end{figure}

To analyze the kinetic energy associated to the defect complex, we calculate 
$E_k$ for the simulation cell with and without the hydrogen molecule:
$E_k$(defect) = $E_k$(64 C + H$_2$) -- $E_k$(64 C), where we use
results obtained in both series of PIMD simulations, with and without
the hydrogen molecule in the graphite cell.
In Fig.~7 we display the kinetic energy as a function of temperature for
H$_2$ (squares), D$_2$ (circles), and T$_2$ (triangles).
As expected, $E_k$ increases as temperature rises, and at a given
temperature, it is larger for smaller isotopic mass. 
At 300 K, we find $E_k$ = 0.238 eV for H$_2$, 0.181 eV for D$_2$,
and 0.154 eV for T$_2$,
which gives isotopic ratios: $E_k$(H$_2$)/$E_k$(D$_2$) = 1.31
and $E_k$(H$_2$)/$E_k$(T$_2$) = 1.55.
These ratios decrease as temperature is raised, and at 900 K they 
amount to 1.16 and 1.24, respectively,
as in the high-temperature (classical) limit they should converge to
unity.  For comparison we also present in Fig.~7 the kinetic energy
corresponding to a classical model with six degrees 
of freedom ($E_k^{\text cl} = 3 k_B T$, dotted line).
For rising temperature,
the classical kinetic energy approaches the results of PIMD
simulations, in particular those corresponding to the heaviest isotope
(tritium), but at 900 K it is still lower than $E_k$(T$_2$)
by 43 meV (about 15\% of the quantum value). 

In the low-temperature limit the quotient 
$E_k$(H$_2$)/$E_k$(D$_2$) is expected to be close to 1.41, as 
obtained in a harmonic approximation.
From our PIMD simulations at 300 K we obtained
a ratio clearly lower than this value, what can
indeed be due to the presence of anharmonicities in the molecular
motion, but more importantly to the excitation of quantum levels  
higher than the ground state at this finite temperature.
This is particularly true for molecular rotation, which is found
to be rather free in the $(x, y)$ plane.  
Something similar can be said from the ratio
$E_k$(H$_2$)/$E_k$(T$_2$) at 300 K, which is clearly lower than
$\sqrt{3}$.

We have also calculated the kinetic energy of the hydrogen molecule
between rigid graphite sheets. The results for $E_k$ obtained in
this case are shown 
in Fig.~7 as filled squares. At 300 K we find $E_k$ = 0.361 eV,
clearly higher than in the case of mobile carbon atoms. 
This decrease in kinetic energy of H$_2$ for flexible graphite sheets
is mainly due to a softening of the vibrational modes corresponding
to motion of the center-of-mass of the molecule in the interlayer
space. In fact, relaxation of the C atoms in the presence of the
H$_2$ molecule causes a decrease in the energy barriers confining the
molecule at a given position, eventually favoring its 
motion in the graphite bulk.
This decrease in $E_k$ is then related to the larger vibrational 
amplitude of the whole molecule between flexible graphite layers,
indicating a nonnegligible coupling in the motion of interstitial
molecule and host atoms.
In connection with this, in an earlier work based on classical
molecular dynamics simulations,\cite{he10} it was shown that 
relaxation of the graphite sheets in the presence of the H$_2$ 
molecule helps to enhance molecular diffusion in the interlayer 
space. In particular, it was found a behavior that
could not be simply explained by a single activation energy, but
suggested the presence of correlations between successive molecular
hops.

\section{Summary}

We have presented results of PIMD simulations for isolated hydrogen 
molecules adsorbed in the interlayer region of graphite.
This kind of simulations allow us to calculate kinetic and potential 
energies at finite temperatures, taking into account the
quantization of host-atom motions, which is not 
easy to consider in fixed-lattice calculations. 
This includes consideration of zero-point motion of guest and
host atoms, which can be coupled in a non-trivial way in the many-body
problem.
Also, isotope effects can be readily explored, since the impurity mass
appears as a parameter in the calculations.

Hydrogen molecules are found to be disposed basically parallel to 
the graphite-layer plane, and free to rotate in this plane.
Although thermal and quantum delocalization allow the molecule to
explore other orientations, molecular rotation is restricted by
the nearest graphite sheets and in fact the H--H axis is not found
to approach the direction perpendicular to the layers.

An important feature of H$_2$ molecules adsorbed in solids 
is their stretching vibration $\omega_s$. 
For molecular hydrogen in graphite, we find at 300 K a frequency
$\omega_s$ = 3916 cm$^{-1}$, to be compared with 
$\omega_s$ = 4055 cm$^{-1}$ obtained for an isolated molecule with
the same interatomic potential and at the same temperature.
For D$_2$ in graphite we find a frequency of 2816 cm$^{-1}$,
which gives an isotopic ratio $\omega_s$(H)/$\omega_s$(D) = 1.39,
similar to those measured for free hydrogen molecules.
It is remarkable that classical simulations yield for H$_2$ a
frequency $\omega_s$ about 160 cm$^{-1}$ larger than the 
PIMD simulations.
The stretching frequency of H$_2$ and D$_2$ is found to decrease
slightly as temperature rises, as a consequence of coupling with
molecular rotation and anharmonicities in the interatomic potential.

 Results of the PIMD simulations including full quantum motion of
all atoms have been compared with those obtained for rigid graphite
layers. This comparison has shown that motion of carbon atoms affects
appreciably several properties of the adsorbed molecules, i.e.,
interatomic distance, stretching frequency, kinetic energy, and atomic
delocalization.

A challenging question, that should be taken into account
in future work, refers to considering coupling between nuclear spins 
in the hydrogen molecule, i.e. dealing separately with ortho and 
para-H$_2$. 
This is particularly important at low temperatures, where
the quantum nature of molecular rotation has to be explicitly
considered in the simulations. 
A quantum treatment of the full problem is not trivial, being mainly
complicated by the ro-vibrational coupling. 
Apart from equilibrium PIMD simulations such as those presented here,
one can apply similar methods to study quantum diffusion of 
H$_2$ in graphite, by calculating free-energy barriers as in
the case of atomic hydrogen in metals\cite{ma95b} and 
semiconductors.\cite{he97,he07}

\begin{acknowledgments}
This work was supported by Ministerio de Ciencia e Innovaci\'on (Spain)
through Grants FIS2006-12117-C04-03 and FIS2009-12721-C04-04,
and by Comunidad Aut\'onoma de Madrid through Program
MODELICO-CM/S2009ESP-1691.
\end{acknowledgments}


\begin{thebibliography}{62}
\expandafter\ifx\csname natexlab\endcsname\relax\def\natexlab#1{#1}\fi
\expandafter\ifx\csname bibnamefont\endcsname\relax
  \def\bibnamefont#1{#1}\fi
\expandafter\ifx\csname bibfnamefont\endcsname\relax
  \def\bibfnamefont#1{#1}\fi
\expandafter\ifx\csname citenamefont\endcsname\relax
  \def\citenamefont#1{#1}\fi
\expandafter\ifx\csname url\endcsname\relax
  \def\url#1{\texttt{#1}}\fi
\expandafter\ifx\csname urlprefix\endcsname\relax\def\urlprefix{URL }\fi
\providecommand{\bibinfo}[2]{#2}
\providecommand{\eprint}[2][]{\url{#2}}

\bibitem[{\citenamefont{Katsnelson}(2007)}]{ka07}
\bibinfo{author}{\bibfnamefont{M.~I.} \bibnamefont{Katsnelson}},
  \bibinfo{journal}{Mater. Today} \textbf{\bibinfo{volume}{10}},
  \bibinfo{pages}{20} (\bibinfo{year}{2007}).

\bibitem[{\citenamefont{Geim and Novoselov}(2007)}]{ge07}
\bibinfo{author}{\bibfnamefont{A.~K.} \bibnamefont{Geim}} \bibnamefont{and}
  \bibinfo{author}{\bibfnamefont{K.~S.} \bibnamefont{Novoselov}},
  \bibinfo{journal}{Nat. Mater.} \textbf{\bibinfo{volume}{6}},
  \bibinfo{pages}{183} (\bibinfo{year}{2007}).

\bibitem[{\citenamefont{Kowalczyk et~al.}(2005)\citenamefont{Kowalczyk, Tanaka,
  Holyst, Kaneko, Ohmori, and Miyamoto}}]{ko05}
\bibinfo{author}{\bibfnamefont{P.}~\bibnamefont{Kowalczyk}},
  \bibinfo{author}{\bibfnamefont{H.}~\bibnamefont{Tanaka}},
  \bibinfo{author}{\bibfnamefont{R.}~\bibnamefont{Holyst}},
  \bibinfo{author}{\bibfnamefont{K.}~\bibnamefont{Kaneko}},
  \bibinfo{author}{\bibfnamefont{T.}~\bibnamefont{Ohmori}}, \bibnamefont{and}
  \bibinfo{author}{\bibfnamefont{J.}~\bibnamefont{Miyamoto}},
  \bibinfo{journal}{J. Phys. Chem. B} \textbf{\bibinfo{volume}{109}},
  \bibinfo{pages}{17174} (\bibinfo{year}{2005}).

\bibitem[{\citenamefont{Dillon and Heben}(2001)}]{di01}
\bibinfo{author}{\bibfnamefont{A.~C.} \bibnamefont{Dillon}} \bibnamefont{and}
  \bibinfo{author}{\bibfnamefont{M.~J.} \bibnamefont{Heben}},
  \bibinfo{journal}{Appl. Phys. A} \textbf{\bibinfo{volume}{72}},
  \bibinfo{pages}{133} (\bibinfo{year}{2001}).

\bibitem[{\citenamefont{Sluiter and Kawazoe}(2003)}]{sl03}
\bibinfo{author}{\bibfnamefont{M.~H.~F.} \bibnamefont{Sluiter}}
  \bibnamefont{and} \bibinfo{author}{\bibfnamefont{Y.}~\bibnamefont{Kawazoe}},
  \bibinfo{journal}{Phys. Rev. B} \textbf{\bibinfo{volume}{68}},
  \bibinfo{pages}{085410} (\bibinfo{year}{2003}).

\bibitem[{\citenamefont{Estreicher}(1995)}]{es95}
\bibinfo{author}{\bibfnamefont{S.~K.} \bibnamefont{Estreicher}},
  \bibinfo{journal}{Mater. Sci. Eng.} \textbf{\bibinfo{volume}{R14}},
  \bibinfo{pages}{319} (\bibinfo{year}{1995}).

\bibitem[{\citenamefont{Pearton et~al.}(1992)\citenamefont{Pearton, Corbett,
  and Stavola}}]{pe92}
\bibinfo{author}{\bibfnamefont{S.~J.} \bibnamefont{Pearton}},
  \bibinfo{author}{\bibfnamefont{J.~W.} \bibnamefont{Corbett}},
  \bibnamefont{and} \bibinfo{author}{\bibfnamefont{M.}~\bibnamefont{Stavola}},
  \emph{\bibinfo{title}{Hydrogen in Crystalline Semiconductors}}
  (\bibinfo{publisher}{Springer}, \bibinfo{address}{Berlin},
  \bibinfo{year}{1992}).

\bibitem[{\citenamefont{Zeisel et~al.}(1999)\citenamefont{Zeisel, Nebel, and
  Stutzmann}}]{ze99}
\bibinfo{author}{\bibfnamefont{R.}~\bibnamefont{Zeisel}},
  \bibinfo{author}{\bibfnamefont{C.~E.} \bibnamefont{Nebel}}, \bibnamefont{and}
  \bibinfo{author}{\bibfnamefont{M.}~\bibnamefont{Stutzmann}},
  \bibinfo{journal}{Appl. Phys. Lett.} \textbf{\bibinfo{volume}{74}},
  \bibinfo{pages}{1875} (\bibinfo{year}{1999}).

\bibitem[{\citenamefont{Atsumi}(2002)}]{at02}
\bibinfo{author}{\bibfnamefont{H.}~\bibnamefont{Atsumi}}, \bibinfo{journal}{J.
  Nucl. Mater.} \textbf{\bibinfo{volume}{307-311}}, \bibinfo{pages}{1466}
  (\bibinfo{year}{2002}).

\bibitem[{\citenamefont{Warrier et~al.}(2007)\citenamefont{Warrier, Schneider,
  Salonen, and Nordlund}}]{wa07}
\bibinfo{author}{\bibfnamefont{M.}~\bibnamefont{Warrier}},
  \bibinfo{author}{\bibfnamefont{R.}~\bibnamefont{Schneider}},
  \bibinfo{author}{\bibfnamefont{E.}~\bibnamefont{Salonen}}, \bibnamefont{and}
  \bibinfo{author}{\bibfnamefont{K.}~\bibnamefont{Nordlund}},
  \bibinfo{journal}{Nucl. Fusion} \textbf{\bibinfo{volume}{47}},
  \bibinfo{pages}{1656} (\bibinfo{year}{2007}).

\bibitem[{\citenamefont{Warrier et~al.}(2004)\citenamefont{Warrier, Schneider,
  Salonen, and Nordlund}}]{wa04}
\bibinfo{author}{\bibfnamefont{M.}~\bibnamefont{Warrier}},
  \bibinfo{author}{\bibfnamefont{R.}~\bibnamefont{Schneider}},
  \bibinfo{author}{\bibfnamefont{E.}~\bibnamefont{Salonen}}, \bibnamefont{and}
  \bibinfo{author}{\bibfnamefont{K.}~\bibnamefont{Nordlund}},
  \bibinfo{journal}{Physica Scripta} \textbf{\bibinfo{volume}{T108}},
  \bibinfo{pages}{85} (\bibinfo{year}{2004}).

\bibitem[{\citenamefont{Shimizu and Tachikawa}(2003)}]{sh03}
\bibinfo{author}{\bibfnamefont{A.}~\bibnamefont{Shimizu}} \bibnamefont{and}
  \bibinfo{author}{\bibfnamefont{H.}~\bibnamefont{Tachikawa}},
  \bibinfo{journal}{J. Phys. Chem. Solids} \textbf{\bibinfo{volume}{64}},
  \bibinfo{pages}{419} (\bibinfo{year}{2003}).

\bibitem[{\citenamefont{Ferro et~al.}(2002)\citenamefont{Ferro, Marinelli, and
  Allouche}}]{fe02}
\bibinfo{author}{\bibfnamefont{Y.}~\bibnamefont{Ferro}},
  \bibinfo{author}{\bibfnamefont{F.}~\bibnamefont{Marinelli}},
  \bibnamefont{and} \bibinfo{author}{\bibfnamefont{A.}~\bibnamefont{Allouche}},
  \bibinfo{journal}{J. Chem. Phys.} \textbf{\bibinfo{volume}{116}},
  \bibinfo{pages}{8124} (\bibinfo{year}{2002}).

\bibitem[{\citenamefont{{Di\~no} et~al.}(2003)\citenamefont{{Di\~no}, Miura,
  Nakanishi, Kasai, and Sugimoto}}]{di03}
\bibinfo{author}{\bibfnamefont{W.~A.} \bibnamefont{{Di\~no}}},
  \bibinfo{author}{\bibfnamefont{Y.}~\bibnamefont{Miura}},
  \bibinfo{author}{\bibfnamefont{H.}~\bibnamefont{Nakanishi}},
  \bibinfo{author}{\bibfnamefont{H.}~\bibnamefont{Kasai}}, \bibnamefont{and}
  \bibinfo{author}{\bibfnamefont{T.}~\bibnamefont{Sugimoto}},
  \bibinfo{journal}{J. Phys. Soc. Japan} \textbf{\bibinfo{volume}{72}},
  \bibinfo{pages}{1867} (\bibinfo{year}{2003}).

\bibitem[{\citenamefont{Ferro et~al.}(2003)\citenamefont{Ferro, Marinelli, and
  Allouche}}]{fe03}
\bibinfo{author}{\bibfnamefont{Y.}~\bibnamefont{Ferro}},
  \bibinfo{author}{\bibfnamefont{F.}~\bibnamefont{Marinelli}},
  \bibnamefont{and} \bibinfo{author}{\bibfnamefont{A.}~\bibnamefont{Allouche}},
  \bibinfo{journal}{Chem. Phys. Lett.} \textbf{\bibinfo{volume}{368}},
  \bibinfo{pages}{609} (\bibinfo{year}{2003}).

\bibitem[{\citenamefont{Ferro et~al.}(2004)\citenamefont{Ferro, Martinelli,
  Jelea, and Allouche}}]{fe04}
\bibinfo{author}{\bibfnamefont{Y.}~\bibnamefont{Ferro}},
  \bibinfo{author}{\bibfnamefont{F.}~\bibnamefont{Martinelli}},
  \bibinfo{author}{\bibfnamefont{A.}~\bibnamefont{Jelea}}, \bibnamefont{and}
  \bibinfo{author}{\bibfnamefont{A.}~\bibnamefont{Allouche}},
  \bibinfo{journal}{J. Chem. Phys.} \textbf{\bibinfo{volume}{120}},
  \bibinfo{pages}{11882} (\bibinfo{year}{2004}).

\bibitem[{\citenamefont{Sha et~al.}(2005)\citenamefont{Sha, Jackson, Lemoine,
  and Lepetit}}]{sh05b}
\bibinfo{author}{\bibfnamefont{X.}~\bibnamefont{Sha}},
  \bibinfo{author}{\bibfnamefont{B.}~\bibnamefont{Jackson}},
  \bibinfo{author}{\bibfnamefont{D.}~\bibnamefont{Lemoine}}, \bibnamefont{and}
  \bibinfo{author}{\bibfnamefont{B.}~\bibnamefont{Lepetit}},
  \bibinfo{journal}{J. Chem. Phys.} \textbf{\bibinfo{volume}{122}},
  \bibinfo{pages}{014709} (\bibinfo{year}{2005}).

\bibitem[{\citenamefont{Morisset and Allouche}(2008)}]{mo08}
\bibinfo{author}{\bibfnamefont{S.}~\bibnamefont{Morisset}} \bibnamefont{and}
  \bibinfo{author}{\bibfnamefont{A.}~\bibnamefont{Allouche}},
  \bibinfo{journal}{J. Chem. Phys.} \textbf{\bibinfo{volume}{129}},
  \bibinfo{pages}{024509} (\bibinfo{year}{2008}).

\bibitem[{\citenamefont{Duplock et~al.}(2004)\citenamefont{Duplock, Scheffler,
  and Lindan}}]{du04}
\bibinfo{author}{\bibfnamefont{E.~J.} \bibnamefont{Duplock}},
  \bibinfo{author}{\bibfnamefont{M.}~\bibnamefont{Scheffler}},
  \bibnamefont{and} \bibinfo{author}{\bibfnamefont{P.~J.~D.}
  \bibnamefont{Lindan}}, \bibinfo{journal}{Phys. Rev. Lett.}
  \textbf{\bibinfo{volume}{92}}, \bibinfo{pages}{225502}
  (\bibinfo{year}{2004}).

\bibitem[{\citenamefont{Yazyev and Helm}(2007)}]{ya07}
\bibinfo{author}{\bibfnamefont{O.~V.} \bibnamefont{Yazyev}} \bibnamefont{and}
  \bibinfo{author}{\bibfnamefont{L.}~\bibnamefont{Helm}},
  \bibinfo{journal}{Phys. Rev. B} \textbf{\bibinfo{volume}{75}},
  \bibinfo{pages}{125408} (\bibinfo{year}{2007}).

\bibitem[{\citenamefont{de~Andres and Verg\'es}(2008)}]{an08}
\bibinfo{author}{\bibfnamefont{P.~L.} \bibnamefont{de~Andres}}
  \bibnamefont{and} \bibinfo{author}{\bibfnamefont{J.~A.}
  \bibnamefont{Verg\'es}}, \bibinfo{journal}{Appl. Phys. Lett.}
  \textbf{\bibinfo{volume}{93}}, \bibinfo{pages}{171915}
  (\bibinfo{year}{2008}).

\bibitem[{\citenamefont{Boukhvalov et~al.}(2008)\citenamefont{Boukhvalov,
  Katsnelson, and Lichtenstein}}]{bo08}
\bibinfo{author}{\bibfnamefont{D.~W.} \bibnamefont{Boukhvalov}},
  \bibinfo{author}{\bibfnamefont{M.~I.} \bibnamefont{Katsnelson}},
  \bibnamefont{and} \bibinfo{author}{\bibfnamefont{A.~I.}
  \bibnamefont{Lichtenstein}}, \bibinfo{journal}{Phys. Rev. B}
  \textbf{\bibinfo{volume}{77}}, \bibinfo{pages}{035427}
  (\bibinfo{year}{2008}).

\bibitem[{\citenamefont{Casolo et~al.}(2009)\citenamefont{Casolo, Lovvik,
  Martinazzo, and Tantardini}}]{ca09}
\bibinfo{author}{\bibfnamefont{S.}~\bibnamefont{Casolo}},
  \bibinfo{author}{\bibfnamefont{O.~M.} \bibnamefont{Lovvik}},
  \bibinfo{author}{\bibfnamefont{R.}~\bibnamefont{Martinazzo}},
  \bibnamefont{and} \bibinfo{author}{\bibfnamefont{G.~F.}
  \bibnamefont{Tantardini}}, \bibinfo{journal}{J. Chem. Phys.}
  \textbf{\bibinfo{volume}{130}}, \bibinfo{pages}{054704}
  (\bibinfo{year}{2009}).

\bibitem[{\citenamefont{Okamoto et~al.}(1997)\citenamefont{Okamoto, Saito, and
  Oshiyama}}]{ok97}
\bibinfo{author}{\bibfnamefont{Y.}~\bibnamefont{Okamoto}},
  \bibinfo{author}{\bibfnamefont{M.}~\bibnamefont{Saito}}, \bibnamefont{and}
  \bibinfo{author}{\bibfnamefont{A.}~\bibnamefont{Oshiyama}},
  \bibinfo{journal}{Phys. Rev. B} \textbf{\bibinfo{volume}{56}},
  \bibinfo{pages}{R10016} (\bibinfo{year}{1997}).

\bibitem[{\citenamefont{Okamoto et~al.}(1998)\citenamefont{Okamoto, Saito, and
  Oshiyama}}]{ok98}
\bibinfo{author}{\bibfnamefont{Y.}~\bibnamefont{Okamoto}},
  \bibinfo{author}{\bibfnamefont{M.}~\bibnamefont{Saito}}, \bibnamefont{and}
  \bibinfo{author}{\bibfnamefont{A.}~\bibnamefont{Oshiyama}},
  \bibinfo{journal}{Phys. Rev. B} \textbf{\bibinfo{volume}{58}},
  \bibinfo{pages}{7701} (\bibinfo{year}{1998}).

\bibitem[{\citenamefont{Hourahine et~al.}(1998)\citenamefont{Hourahine, Jones,
  \"Oberg, Newman, Briddon, and Roduner}}]{ho98}
\bibinfo{author}{\bibfnamefont{B.}~\bibnamefont{Hourahine}},
  \bibinfo{author}{\bibfnamefont{R.}~\bibnamefont{Jones}},
  \bibinfo{author}{\bibfnamefont{S.}~\bibnamefont{\"Oberg}},
  \bibinfo{author}{\bibfnamefont{R.~C.} \bibnamefont{Newman}},
  \bibinfo{author}{\bibfnamefont{P.~R.} \bibnamefont{Briddon}},
  \bibnamefont{and} \bibinfo{author}{\bibfnamefont{E.}~\bibnamefont{Roduner}},
  \bibinfo{journal}{Phys. Rev. B} \textbf{\bibinfo{volume}{57}},
  \bibinfo{pages}{R12666} (\bibinfo{year}{1998}).

\bibitem[{\citenamefont{{Van~de~Walle}}(1998)}]{wa98}
\bibinfo{author}{\bibfnamefont{C.~G.} \bibnamefont{{Van~de~Walle}}},
  \bibinfo{journal}{Phys. Rev. Lett.} \textbf{\bibinfo{volume}{80}},
  \bibinfo{pages}{2177} (\bibinfo{year}{1998}).

\bibitem[{\citenamefont{Pruneda et~al.}(2002)\citenamefont{Pruneda, Estreicher,
  Junquera, Ferrer, and Ordej\'on}}]{pr02}
\bibinfo{author}{\bibfnamefont{J.~M.} \bibnamefont{Pruneda}},
  \bibinfo{author}{\bibfnamefont{S.~K.} \bibnamefont{Estreicher}},
  \bibinfo{author}{\bibfnamefont{J.}~\bibnamefont{Junquera}},
  \bibinfo{author}{\bibfnamefont{J.}~\bibnamefont{Ferrer}}, \bibnamefont{and}
  \bibinfo{author}{\bibfnamefont{P.}~\bibnamefont{Ordej\'on}},
  \bibinfo{journal}{Phys. Rev. B} \textbf{\bibinfo{volume}{65}},
  \bibinfo{pages}{075210} (\bibinfo{year}{2002}).

\bibitem[{\citenamefont{Fowler et~al.}(2002)\citenamefont{Fowler, Walters, and
  Stavola}}]{fo02}
\bibinfo{author}{\bibfnamefont{W.~B.} \bibnamefont{Fowler}},
  \bibinfo{author}{\bibfnamefont{P.}~\bibnamefont{Walters}}, \bibnamefont{and}
  \bibinfo{author}{\bibfnamefont{M.}~\bibnamefont{Stavola}},
  \bibinfo{journal}{Phys. Rev. B} \textbf{\bibinfo{volume}{66}},
  \bibinfo{pages}{075216} (\bibinfo{year}{2002}).

\bibitem[{\citenamefont{Hourahine and Jones}(2003)}]{ho03}
\bibinfo{author}{\bibfnamefont{B.}~\bibnamefont{Hourahine}} \bibnamefont{and}
  \bibinfo{author}{\bibfnamefont{R.}~\bibnamefont{Jones}},
  \bibinfo{journal}{Phys. Rev. B} \textbf{\bibinfo{volume}{67}},
  \bibinfo{pages}{121205(R)} (\bibinfo{year}{2003}).

\bibitem[{\citenamefont{Gillan}(1988)}]{gi88}
\bibinfo{author}{\bibfnamefont{M.~J.} \bibnamefont{Gillan}},
  \bibinfo{journal}{Phil. Mag. A} \textbf{\bibinfo{volume}{58}},
  \bibinfo{pages}{257} (\bibinfo{year}{1988}).

\bibitem[{\citenamefont{Ceperley}(1995)}]{ce95}
\bibinfo{author}{\bibfnamefont{D.~M.} \bibnamefont{Ceperley}},
  \bibinfo{journal}{Rev. Mod. Phys.} \textbf{\bibinfo{volume}{67}},
  \bibinfo{pages}{279} (\bibinfo{year}{1995}).

\bibitem[{\citenamefont{Ram\'{\i}rez and Herrero}(1994)}]{ra94}
\bibinfo{author}{\bibfnamefont{R.}~\bibnamefont{Ram\'{\i}rez}}
  \bibnamefont{and} \bibinfo{author}{\bibfnamefont{C.~P.}
  \bibnamefont{Herrero}}, \bibinfo{journal}{Phys. Rev. Lett.}
  \textbf{\bibinfo{volume}{73}}, \bibinfo{pages}{126} (\bibinfo{year}{1994}).

\bibitem[{\citenamefont{Herrero and Ram\'{\i}rez}(1995)}]{he95}
\bibinfo{author}{\bibfnamefont{C.~P.} \bibnamefont{Herrero}} \bibnamefont{and}
  \bibinfo{author}{\bibfnamefont{R.}~\bibnamefont{Ram\'{\i}rez}},
  \bibinfo{journal}{Phys. Rev. B} \textbf{\bibinfo{volume}{51}},
  \bibinfo{pages}{16761} (\bibinfo{year}{1995}).

\bibitem[{\citenamefont{Miyake et~al.}(1998)\citenamefont{Miyake, Ogitsu, and
  Tsuneyuki}}]{mi98}
\bibinfo{author}{\bibfnamefont{T.}~\bibnamefont{Miyake}},
  \bibinfo{author}{\bibfnamefont{T.}~\bibnamefont{Ogitsu}}, \bibnamefont{and}
  \bibinfo{author}{\bibfnamefont{S.}~\bibnamefont{Tsuneyuki}},
  \bibinfo{journal}{Phys. Rev. Lett.} \textbf{\bibinfo{volume}{81}},
  \bibinfo{pages}{1873} (\bibinfo{year}{1998}).

\bibitem[{\citenamefont{Herrero et~al.}(2006)\citenamefont{Herrero,
  Ram\'{\i}rez, and Hern\'andez}}]{he06}
\bibinfo{author}{\bibfnamefont{C.~P.} \bibnamefont{Herrero}},
  \bibinfo{author}{\bibfnamefont{R.}~\bibnamefont{Ram\'{\i}rez}},
  \bibnamefont{and} \bibinfo{author}{\bibfnamefont{E.~R.}
  \bibnamefont{Hern\'andez}}, \bibinfo{journal}{Phys. Rev. B}
  \textbf{\bibinfo{volume}{73}}, \bibinfo{pages}{245211}
  (\bibinfo{year}{2006}).

\bibitem[{\citenamefont{Herrero and Ram\'{\i}rez}(2007)}]{he07}
\bibinfo{author}{\bibfnamefont{C.~P.} \bibnamefont{Herrero}} \bibnamefont{and}
  \bibinfo{author}{\bibfnamefont{R.}~\bibnamefont{Ram\'{\i}rez}},
  \bibinfo{journal}{Phys. Rev. Lett.} \textbf{\bibinfo{volume}{99}},
  \bibinfo{pages}{205504} (\bibinfo{year}{2007}).

\bibitem[{\citenamefont{Mattsson and Wahnstr\"om}(1995)}]{ma95b}
\bibinfo{author}{\bibfnamefont{T.~R.} \bibnamefont{Mattsson}} \bibnamefont{and}
  \bibinfo{author}{\bibfnamefont{G.}~\bibnamefont{Wahnstr\"om}},
  \bibinfo{journal}{Phys. Rev. B} \textbf{\bibinfo{volume}{51}},
  \bibinfo{pages}{1885} (\bibinfo{year}{1995}).

\bibitem[{\citenamefont{Herrero and Ram\'{\i}rez}(2009{\natexlab{a}})}]{he09a}
\bibinfo{author}{\bibfnamefont{C.~P.} \bibnamefont{Herrero}} \bibnamefont{and}
  \bibinfo{author}{\bibfnamefont{R.}~\bibnamefont{Ram\'{\i}rez}},
  \bibinfo{journal}{Phys. Rev. B} \textbf{\bibinfo{volume}{79}},
  \bibinfo{pages}{115429} (\bibinfo{year}{2009}{\natexlab{a}}).

\bibitem[{\citenamefont{Gordillo et~al.}(2001)\citenamefont{Gordillo, Boronat,
  and Casulleras}}]{go01}
\bibinfo{author}{\bibfnamefont{M.~C.} \bibnamefont{Gordillo}},
  \bibinfo{author}{\bibfnamefont{J.}~\bibnamefont{Boronat}}, \bibnamefont{and}
  \bibinfo{author}{\bibfnamefont{J.}~\bibnamefont{Casulleras}},
  \bibinfo{journal}{Phys. Rev. B} \textbf{\bibinfo{volume}{65}},
  \bibinfo{pages}{014503} (\bibinfo{year}{2001}).

\bibitem[{\citenamefont{Chakravarty}(1999)}]{ch99}
\bibinfo{author}{\bibfnamefont{C.}~\bibnamefont{Chakravarty}},
  \bibinfo{journal}{Phys. Rev. B} \textbf{\bibinfo{volume}{59}},
  \bibinfo{pages}{3590} (\bibinfo{year}{1999}).

\bibitem[{\citenamefont{Kaxiras and Guo}(1994)}]{ka94}
\bibinfo{author}{\bibfnamefont{E.}~\bibnamefont{Kaxiras}} \bibnamefont{and}
  \bibinfo{author}{\bibfnamefont{Z.}~\bibnamefont{Guo}},
  \bibinfo{journal}{Phys. Rev. B} \textbf{\bibinfo{volume}{49}},
  \bibinfo{pages}{11822} (\bibinfo{year}{1994}).

\bibitem[{\citenamefont{Surh et~al.}(1997)\citenamefont{Surh, Runge, Barbee,
  Pollock, and Mailhiot}}]{su97}
\bibinfo{author}{\bibfnamefont{M.~P.} \bibnamefont{Surh}},
  \bibinfo{author}{\bibfnamefont{K.~J.} \bibnamefont{Runge}},
  \bibinfo{author}{\bibfnamefont{T.~W.} \bibnamefont{Barbee}},
  \bibinfo{author}{\bibfnamefont{E.~L.} \bibnamefont{Pollock}},
  \bibnamefont{and} \bibinfo{author}{\bibfnamefont{C.}~\bibnamefont{Mailhiot}},
  \bibinfo{journal}{Phys. Rev. B} \textbf{\bibinfo{volume}{55}},
  \bibinfo{pages}{11330} (\bibinfo{year}{1997}).

\bibitem[{\citenamefont{Kitamura et~al.}(2000)\citenamefont{Kitamura,
  Tsuneyuki, Ogitsu, and Miyake}}]{ki00}
\bibinfo{author}{\bibfnamefont{H.}~\bibnamefont{Kitamura}},
  \bibinfo{author}{\bibfnamefont{S.}~\bibnamefont{Tsuneyuki}},
  \bibinfo{author}{\bibfnamefont{T.}~\bibnamefont{Ogitsu}}, \bibnamefont{and}
  \bibinfo{author}{\bibfnamefont{T.}~\bibnamefont{Miyake}},
  \bibinfo{journal}{Nature} \textbf{\bibinfo{volume}{404}},
  \bibinfo{pages}{259} (\bibinfo{year}{2000}).

\bibitem[{\citenamefont{Feynman}(1972)}]{fe72}
\bibinfo{author}{\bibfnamefont{R.~P.} \bibnamefont{Feynman}},
  \emph{\bibinfo{title}{Statistical Mechanics}}
  (\bibinfo{publisher}{Addison-Wesley}, \bibinfo{address}{New York},
  \bibinfo{year}{1972}).

\bibitem[{\citenamefont{Kleinert}(1990)}]{kl90}
\bibinfo{author}{\bibfnamefont{H.}~\bibnamefont{Kleinert}},
  \emph{\bibinfo{title}{Path Integrals in Quantum Mechanics, Statistics and
  Polymer Physics}} (\bibinfo{publisher}{World Scientific},
  \bibinfo{address}{Singapore}, \bibinfo{year}{1990}).

\bibitem[{\citenamefont{Martyna et~al.}(1996)\citenamefont{Martyna, Tuckerman,
  Tobias, and Klein}}]{ma96}
\bibinfo{author}{\bibfnamefont{G.~J.} \bibnamefont{Martyna}},
  \bibinfo{author}{\bibfnamefont{M.~E.} \bibnamefont{Tuckerman}},
  \bibinfo{author}{\bibfnamefont{D.~J.} \bibnamefont{Tobias}},
  \bibnamefont{and} \bibinfo{author}{\bibfnamefont{M.~L.} \bibnamefont{Klein}},
  \bibinfo{journal}{Mol. Phys.} \textbf{\bibinfo{volume}{87}},
  \bibinfo{pages}{1117} (\bibinfo{year}{1996}).

\bibitem[{\citenamefont{Tuckerman}(2002)}]{tu02}
\bibinfo{author}{\bibfnamefont{M.~E.} \bibnamefont{Tuckerman}}, in
  \emph{\bibinfo{booktitle}{Quantum Simulations of Complex Many--Body Systems:
  From Theory to Algorithms}}, edited by
  \bibinfo{editor}{\bibfnamefont{J.}~\bibnamefont{Grotendorst}},
  \bibinfo{editor}{\bibfnamefont{D.}~\bibnamefont{Marx}}, \bibnamefont{and}
  \bibinfo{editor}{\bibfnamefont{A.}~\bibnamefont{Muramatsu}}
  (\bibinfo{publisher}{NIC}, \bibinfo{address}{FZ J\"ulich},
  \bibinfo{year}{2002}), p. \bibinfo{pages}{269}.

\bibitem[{\citenamefont{Porezag et~al.}(1995)\citenamefont{Porezag, Frauenheim,
  K\"ohler, Seifert, and Kaschner}}]{po95}
\bibinfo{author}{\bibfnamefont{D.}~\bibnamefont{Porezag}},
  \bibinfo{author}{\bibfnamefont{T.}~\bibnamefont{Frauenheim}},
  \bibinfo{author}{\bibfnamefont{T.}~\bibnamefont{K\"ohler}},
  \bibinfo{author}{\bibfnamefont{G.}~\bibnamefont{Seifert}}, \bibnamefont{and}
  \bibinfo{author}{\bibfnamefont{R.}~\bibnamefont{Kaschner}},
  \bibinfo{journal}{Phys. Rev. B} \textbf{\bibinfo{volume}{51}},
  \bibinfo{pages}{12947} (\bibinfo{year}{1995}).

\bibitem[{\citenamefont{Goringe et~al.}(1997)\citenamefont{Goringe, Bowler, and
  Hern\'andez}}]{go97}
\bibinfo{author}{\bibfnamefont{C.~M.} \bibnamefont{Goringe}},
  \bibinfo{author}{\bibfnamefont{D.~R.} \bibnamefont{Bowler}},
  \bibnamefont{and}
  \bibinfo{author}{\bibfnamefont{E.}~\bibnamefont{Hern\'andez}},
  \bibinfo{journal}{Rep. Prog. Phys.} \textbf{\bibinfo{volume}{60}},
  \bibinfo{pages}{1447} (\bibinfo{year}{1997}).

\bibitem[{\citenamefont{L{\'o}pez-Ciudad
  et~al.}(2003)\citenamefont{L{\'o}pez-Ciudad, Ram{\'{\i}}rez, Schulte, and
  B\"ohm}}]{lo03}
\bibinfo{author}{\bibfnamefont{T.}~\bibnamefont{L{\'o}pez-Ciudad}},
  \bibinfo{author}{\bibfnamefont{R.}~\bibnamefont{Ram{\'{\i}}rez}},
  \bibinfo{author}{\bibfnamefont{J.}~\bibnamefont{Schulte}}, \bibnamefont{and}
  \bibinfo{author}{\bibfnamefont{M.~C.} \bibnamefont{B\"ohm}},
  \bibinfo{journal}{J. Chem. Phys.} \textbf{\bibinfo{volume}{119}},
  \bibinfo{pages}{4328} (\bibinfo{year}{2003}).

\bibitem[{\citenamefont{B\"ohm et~al.}(2001)\citenamefont{B\"ohm, Schulte,
  Hern\'andez, and Ram{\'{\i}}rez}}]{bo01}
\bibinfo{author}{\bibfnamefont{M.~C.} \bibnamefont{B\"ohm}},
  \bibinfo{author}{\bibfnamefont{J.}~\bibnamefont{Schulte}},
  \bibinfo{author}{\bibfnamefont{E.}~\bibnamefont{Hern\'andez}},
  \bibnamefont{and}
  \bibinfo{author}{\bibfnamefont{R.}~\bibnamefont{Ram{\'{\i}}rez}},
  \bibinfo{journal}{Chem. Phys.} \textbf{\bibinfo{volume}{264}},
  \bibinfo{pages}{371} (\bibinfo{year}{2001}).

\bibitem[{\citenamefont{Herrero and Ram\'{\i}rez}(2009{\natexlab{b}})}]{he09b}
\bibinfo{author}{\bibfnamefont{C.~P.} \bibnamefont{Herrero}} \bibnamefont{and}
  \bibinfo{author}{\bibfnamefont{R.}~\bibnamefont{Ram\'{\i}rez}},
  \bibinfo{journal}{Phys. Rev. B} \textbf{\bibinfo{volume}{80}},
  \bibinfo{pages}{035207} (\bibinfo{year}{2009}{\natexlab{b}}).

\bibitem[{\citenamefont{Ram{\'{\i}}rez and L{\'o}pez-Ciudad}(2001)}]{ra01}
\bibinfo{author}{\bibfnamefont{R.}~\bibnamefont{Ram{\'{\i}}rez}}
  \bibnamefont{and}
  \bibinfo{author}{\bibfnamefont{T.}~\bibnamefont{L{\'o}pez-Ciudad}},
  \bibinfo{journal}{J. Chem. Phys.} \textbf{\bibinfo{volume}{115}},
  \bibinfo{pages}{103} (\bibinfo{year}{2001}).

\bibitem[{\citenamefont{Tuckerman and Hughes}(1998)}]{tu98}
\bibinfo{author}{\bibfnamefont{M.~E.} \bibnamefont{Tuckerman}}
  \bibnamefont{and} \bibinfo{author}{\bibfnamefont{A.}~\bibnamefont{Hughes}},
  in \emph{\bibinfo{booktitle}{Classical and Quantum Dynamics in Condensed
  Phase Simulations}}, edited by \bibinfo{editor}{\bibfnamefont{B.~J.}
  \bibnamefont{Berne}},
  \bibinfo{editor}{\bibfnamefont{G.}~\bibnamefont{Ciccotti}}, \bibnamefont{and}
  \bibinfo{editor}{\bibfnamefont{D.~F.} \bibnamefont{Coker}}
  (\bibinfo{publisher}{Word Scientific}, \bibinfo{address}{Singapore},
  \bibinfo{year}{1998}), p. \bibinfo{pages}{311}.

\bibitem[{\citenamefont{Ram\'{\i}rez and L\'opez-Ciudad}(2002)}]{ra02}
\bibinfo{author}{\bibfnamefont{R.}~\bibnamefont{Ram\'{\i}rez}}
  \bibnamefont{and}
  \bibinfo{author}{\bibfnamefont{T.}~\bibnamefont{L\'opez-Ciudad}}, in
  \emph{\bibinfo{booktitle}{Quantum Simulations of Complex Many--Body Systems:
  From Theory to Algorithms}}, edited by
  \bibinfo{editor}{\bibfnamefont{J.}~\bibnamefont{Grotendorst}},
  \bibinfo{editor}{\bibfnamefont{D.}~\bibnamefont{Marx}}, \bibnamefont{and}
  \bibinfo{editor}{\bibfnamefont{A.}~\bibnamefont{Muramatsu}}
  (\bibinfo{publisher}{NIC}, \bibinfo{address}{FZ J\"ulich},
  \bibinfo{year}{2002}), pp. \bibinfo{pages}{325--375; for downloads and
  audio--visual Lecture Notes see {\tt www.theochem.rub.de/go/cprev.html}.}

\bibitem[{\citenamefont{Ram\'{\i}rez and Herrero}(2005)}]{ra05}
\bibinfo{author}{\bibfnamefont{R.}~\bibnamefont{Ram\'{\i}rez}}
  \bibnamefont{and} \bibinfo{author}{\bibfnamefont{C.~P.}
  \bibnamefont{Herrero}}, \bibinfo{journal}{Phys. Rev. B}
  \textbf{\bibinfo{volume}{72}}, \bibinfo{pages}{024303}
  (\bibinfo{year}{2005}).

\bibitem[{\citenamefont{Ram\'{\i}rez and Herrero}(1993)}]{ra93}
\bibinfo{author}{\bibfnamefont{R.}~\bibnamefont{Ram\'{\i}rez}}
  \bibnamefont{and} \bibinfo{author}{\bibfnamefont{C.~P.}
  \bibnamefont{Herrero}}, \bibinfo{journal}{Phys. Rev. B}
  \textbf{\bibinfo{volume}{48}}, \bibinfo{pages}{14659} (\bibinfo{year}{1993}).

\bibitem[{\citenamefont{Herrero and Ram\'{\i}rez}(2010)}]{he10}
\bibinfo{author}{\bibfnamefont{C.~P.} \bibnamefont{Herrero}} \bibnamefont{and}
  \bibinfo{author}{\bibfnamefont{R.}~\bibnamefont{Ram\'{\i}rez}},
  \bibinfo{journal}{J. Phys. D: Appl. Phys.} \textbf{\bibinfo{volume}{43}},
  \bibinfo{pages}{255402} (\bibinfo{year}{2010}).

\bibitem[{\citenamefont{Stoicheff}(1957)}]{st57}
\bibinfo{author}{\bibfnamefont{B.~P.} \bibnamefont{Stoicheff}},
  \bibinfo{journal}{Can. J. Phys.} \textbf{\bibinfo{volume}{35}},
  \bibinfo{pages}{730} (\bibinfo{year}{1957}).

\bibitem[{\citenamefont{Herman et~al.}(1982)\citenamefont{Herman, Bruskin, and
  Berne}}]{he82}
\bibinfo{author}{\bibfnamefont{M.~F.} \bibnamefont{Herman}},
  \bibinfo{author}{\bibfnamefont{E.~J.} \bibnamefont{Bruskin}},
  \bibnamefont{and} \bibinfo{author}{\bibfnamefont{B.~J.} \bibnamefont{Berne}},
  \bibinfo{journal}{J. Chem. Phys.} \textbf{\bibinfo{volume}{76}},
  \bibinfo{pages}{5150} (\bibinfo{year}{1982}).

\bibitem[{\citenamefont{Herrero}(1997)}]{he97}
\bibinfo{author}{\bibfnamefont{C.~P.} \bibnamefont{Herrero}},
  \bibinfo{journal}{Phys. Rev. B} \textbf{\bibinfo{volume}{55}},
  \bibinfo{pages}{9235} (\bibinfo{year}{1997}).

\end{thebibliography}
\end{document}